\documentclass[pra,twocolumn,showpacs,superscriptaddress,floatfix]{revtex4-2}
\usepackage{graphicx}
\usepackage{epstopdf} 
\usepackage{amsmath}
\usepackage{epsfig}
\usepackage{latexsym}
\usepackage{amsfonts}
\usepackage{graphics}
\usepackage{bm}
\usepackage{braket}
\usepackage{dsfont}
\usepackage{soul}
\usepackage{ulem}
\usepackage{bbm}

\usepackage{mathtools}

\usepackage{helvet}

\graphicspath{ {./Graphics/} }
\begin{document}

\date{\today}
\author{J. Mumford}
\affiliation{Department of Physics and Astronomy, McMaster University, 1280 Main St.\ W., Hamilton, ON, L8S 4M1, Canada}

\title{Synthetic gauge field in two interacting ultracold atomic gases without an optical lattice}

\begin{abstract}
A 2D Fock-state lattice (FSL) is constructed from the many-body states of two interacting two-mode quantum gases.  By periodically driving the interspecies interactions and pulsing the tunneling between the two modes of each gas, a synthetic gauge field is generated.  We derive an effective Hamiltonian in the short pulse limit which resembles the Harper-Hofstadter Hamiltonian where the magnetic flux per plaquette is controlled by the ratio of the interaction energy and the driving frequency.  The quasispectrum of the Floquet operator of the driving sequence shows the celebrated Hofstadter's butterfly pattern as well as the existence of edge states.  From the calculation of the local Chern marker, we establish that the FSL has non-trivial topology and by simulating the dynamics of the edge states, show that they exhibit chirality.  Finally, the inclusion of intraspecies interactions creates an overall harmonic trap in the lattice and introduces the nonlinear effect of macroscopic quantum self-trapping which is shown to hinder movement along the edges of the lattice.  This work introduces a new avenue to explore synthetic gauge fields and the link between condensed matter systems and quantum gases.
\end{abstract}

\maketitle

\section{\label{Sec:Intro}Introduction}

Ultracold atomic gases captured in optical lattices have become a versatile testbed for simulating condensed matter systems \cite{jaksch05}.  One of the main methods behind these simulations, called Floquet engineering, involves periodically driving external fields of the gas and lattice to create synthetic gauge fields \cite{dalibard11,goldman14a,goldman14b,burkov15,eckardt15,creffield16}.  The driving can be finely tuned in order to generate an effective Hamiltonian that has the desired properties of a static one.  This is especially useful when simulating the interactions between electrons and applied magnetic fields in materials. Since ultracold gases are kept in clean, highly controlled environments, the synthetic magnetic fields produced in these systems are also highly controlled. Depending on the specific driving schemes used, the parameters of the system can be tuned to generate magnetic field strengths that are difficult to approach otherwise.  An early success in simulating a magnetic field involved the Coriolis force in a rotating atomic gas \cite{madison00,schweikhard04}.  Later, synthetic staggered magnetic fields were created with laser induced tunneling \cite{aidelsburger11} and shaking of the lattice \cite{struck11}.  This lead to the creation of topological many-body phases in 2D optical lattices through the experimental simulation of the Harper-Hofstadter (HH) \cite{aidelsburger13,miyake13} and Haldane \cite{jotzu14} models.  

An additional aid in the engineering of condensed matter systems is the concept of synthetic dimensions \cite{boada12,ozawa19b}.  A synthetic dimension is a degree of freedom, from either internal or external states of particles, which can imitate real-space dimensions.  Experimentalists have been able to create synthetic dimensions using internal spin \cite{mancini15,stuhl15,celi14,anisimovas16}, momentum \cite{an17,meier16,xie19}, clock \cite{wall16,livi16,kolkowitz17}, harmonic oscillator \cite{price17}, and rotational \cite{flob15} states of atoms and molecules.  It is common practice to prepare a 1D lattice in real-space and pair it with a synthetic dimension to create a 2D lattice.  The inclusion of a synthetic gauge field in such systems has lead to the observation of chiral edge states \cite{mancini15,stuhl15} and Hall drift, as well as the construction of the local Chern marker \cite{chalopin20}.

Another avenue taken to generate topological effects is through photonic systems \cite{ozawa19a,smirnova20}.  Usually waveguides and resonators are used as lattice sites arranged in the desired geometry.  In these systems, topological edge states \cite{hafezi13,mukherjee17}, band structures \cite{rechtsman13} and transport properties \cite{mittal16} have been observed.  A relatively newer approach is to use a few photonic modes in the form of optical cavities.  The range of photon occupations of these modes form a Fock-state basis which is used as a synthetic dimension and can be generalized to show that $d$ cavities form a $d-1$ dimensional Fock-state lattice (FSL) \cite{weiwang16,cai21}.  There is some restriction to this approach, however, as the geometry of the FSL is ingrained in the Hilbert space making it difficult to manipulate.  Nevertheless, theoretical treatments of various lattice models with gauge fields have been proposed using FSLs \cite{saugmann22}.

Fock states also form the basis of few-mode matter-based systems such as a two-component Bose-Einstein condensate (BEC) \cite{zibold10} or a BEC in a double well potential, alternatively called a bosonic Josephson junction (BJJ) \cite{smerzi97,andrews97}.  There are three main ways in which these systems are periodically driven: the chemical potential, the tunneling between modes, and the interactions between particles of the gas.  For instance, periodically driving the tilt between two wells of a BJJ has been proposed to enhance tunneling \cite{weiss06} and periodically driving the tunneling between two modes of a Bose-Hubbard (BH) dimer has been used to explore the emergence of chaos in the system \cite{kidd19}.  Recently, periodically driven interactions have been proposed to control the quantum collapse in a BJJ \cite{vera13}, to generate quantum phase transitions in the Lipkin-Meshkov-Glick model \cite{engelhardt13} and to control the tunneling of ultracold atoms in the BH dimer \cite{watanabe12}.

In this work, we show a 2D FSL lattice can be formed by the states of two interacting two-mode quantum gases.  By incorporating a periodic driving sequence used on a real-space lattice to generate synthetic gauge fields \cite{sorensen05}, we show the same can be done for the FSL.  The common thread throughout this paper is the adoption of real-space measures of topology and their conversion to equivalent measures of the FSL.  Consequently, we establish that the FSL is HH-like with non-trivial topology giving rise to nonzero Chern numbers and chiral edge states.  Nonlinearities from the repulsive intraspecies interactions of the gases play an important role in the static and dynamic properties of the system.  They act as an overall harmonic potential in the FSL causing some merging of bulk bands and they manifest in the dynamics in the form of self-trapping, hindering the movement around the edge of the lattice.

\section{\label{Sec:Model}Model}

The system we will be investigating consists of two interacting bosonic quantum gases.  Each gas contains $N$ identical particles and each particle has access to two states labeled 1 and 2.  The states can be internal spin states or external states such as the ground states in each well of a double well potential.  The basis we will be using is the Fock basis defined in terms of half the particle number difference between states 1 and 2, $\vert n, m\rangle$, where $n = \left ( N_1 - N_2 \right )/2$ and $m = \left ( M_1 -M_2 \right )/2$.  In the Schwinger representation, $n$ and $m$ are interpreted as $z$ - components of spin-$N/2$ particles, so $\vert n, m \rangle$ is an eigenstate of the operators $\hat{J}_z = \frac{1}{2} \left ( \hat{a}_1^\dagger\hat{a}_1 - \hat{a}_2^\dagger \hat{a}_2 \right )$ and $\hat{S}_z = \frac{1}{2} \left (\hat{b}_1^\dagger \hat{b}_1 - \hat{b}_2^\dagger \hat{b}_2 \right)$ where $\hat{J}_z \vert n, m \rangle  = n \vert n,m \rangle$ and $\hat{S}_z \vert n,m\rangle = m \vert n,m \rangle$.  The creation and annihilation operators for the two states of the two gases follow the usual commutation relation $[ \hat{a}_i, \hat{a}^\dagger_j  ] =  [ \hat{b}_i, \hat{b}^\dagger_j  ] = \delta_{ij}$ where $i,j = 1,2$. Here, we will interpret the Fock states as locations on a $\left (N+1 \right ) \times \left (N+1 \right )$ FSL with lattice spacing $a = 1$ where $n$ and $m$ are the coordinates along the $x$ and $y$ axis, respectively.  The origin located at $\vert 0,0\rangle$ corresponds to an equal number of particles of each gas occupying each of the two modes and $\hat{J}_z$ ($\hat{S}_z$) becomes the displacement operator $\hat{x}$ ($\hat{y}$) on the FSL.

Our goal in this paper is to investigate the effects of a synthetic magnetic field through the FSL.  To generate the magnetic field, we are motivated by past efforts which have generated one in real-space.  For instance, in \cite{sorensen05} it was shown that a magnetic flux through a real-space lattice could be generated from a combination of periodically driving a quadrupolar potential term $V_{\mathrm{qp}} \sin (\omega t) \hat{x} \hat{y}$, where $\omega$ is the driving frequency, and periodically flashing on tunneling in the $x$ and $y$ directions for a short period of time, $\tau \ll 1$.  One period of the driving sequence produces the Floquet operator

\begin{eqnarray}
\hat{U}^\prime\left (t = \frac{2\pi}{\omega} \right ) = &\mathrm{e}&^{-\mathrm{i} \tau \hat{T}_y/2}  \mathrm{e}^{2 \mathrm{i} V_{\mathrm{qp}}\hat{x}\hat{y}/\omega} \mathrm{e}^{-\mathrm{i} \tau \hat{T}_x}  \mathrm{e}^{-2 \mathrm{i} V_{\mathrm{qp}}\hat{x}\hat{y}/\omega} \nonumber \\
&& \times  \mathrm{e}^{-\mathrm{i} \tau \hat{T}_y/2}
\label{eq:U1}
\end{eqnarray}    
where $\hat{T}_x$ and $\hat{T}_y$ are the kinetic energy operators in the $x$ and $y$ directions, respectively.  For $\omega \gg 1$, the evolution operator can be written in terms of an effective Hamiltonian $\hat{U}^\prime(\tau) = \mathrm{e}^{-\mathrm{i} \hat{H}^\prime_\mathrm{eff}\tau}$ where 

\begin{eqnarray}
\hat{H}_\mathrm{eff}^\prime \approx \hat{H}_\mathrm{HH} = &-&J \sum_{n,m}  \vert n,m \rangle \langle n,m-1 \vert \nonumber \\
&& + \mathrm{e}^{\mathrm{i} 2\pi \alpha m} \vert n, m \rangle \langle n-1, m\vert + \mathrm{h.c} \, .
\label{eq:harper}
\end{eqnarray}
The effective Hamiltonian takes the form of the celebrated HH Hamiltonian $\hat{H}_\mathrm{HH}$ (in the Landau gauge) which describes a uniform magnetic field in the direction perpendicular to the plane of a 2D lattice.  The parameter $\alpha = V_\mathrm{qp}/\pi\omega = \Phi/\Phi_0$ is the ratio of the magnetic flux through a unit cell $\Phi = Ba^2$ and the magnetic flux quantum $\Phi_0 = h/e$.

Switching the focus back to our model, according to our previous analogy between $\hat{J}_z$ ($\hat{S}_z$) and $\hat{x}$ ($\hat{y}$), the quadrupolar potential becomes an interaction term between the two gases $V_\mathrm{qp} \hat{x}\hat{y} \to \kappa \hat{S}_z \hat{J}_z$ where $\kappa$ is the interaction energy.  The kinetic energy operators in Eq.\ \eqref{eq:U1} are responsible for moving a particle from one lattice site to an adjacent one.  The $x$-component spin operators accomplish the same thing on the FSL, so we make the additional transformations $\hat{T}_x \to -J\hat{J}_x = -\frac{J}{2} \left (\hat{a}_1^\dagger \hat{a}_2 + \hat{a}^\dagger_2 \hat{a}_1 \right )$ and $\hat{T}_y \to -J\hat{S}_x = -\frac{J}{2} \left ( \hat{b}_1^\dagger \hat{b}_2 + \hat{b}_2^\dagger \hat{b}_1 \right )$ (keeping in mind that the subscripts on the kinetic energy operators represent direction and the subscripts on the spin operators do not) where $J$ is the tunneling energy and is the same for both gases.  The driving sequence in Eq.\ \eqref{eq:U1} becomes

\begin{equation}
\hat{U}\left (t = \frac{2\pi}{\omega} \right ) =  \mathrm{e}^{2 \mathrm{i} \kappa \hat{S}_z\hat{J}_z/\omega} \mathrm{e}^{\mathrm{i} \tau \hat{J}_x}  \mathrm{e}^{-2 \mathrm{i} \kappa \hat{S}_z\hat{J}_z/\omega} \mathrm{e}^{\mathrm{i} \tau \hat{S}_x}
\label{eq:U2}
\end{equation}  
where, when comparing it to Eq.\ \eqref{eq:U1}, we have made the small modification of removing the left-hand factor of $\mathrm{e}^{\mathrm{i} \tau \hat{S}_x/2}$ and multiplying it to the right-hand side.  Also, we have set the tunneling energy to unity, $J=1$, so that from now on all other energies will be in units of $J$.  A single species version of $\hat{U}$ has been used to produce the fractal spectrum of Hofstadter's butterfly \cite{wang10} and study the effects of nonlinearities on it.  This driving sequence is accomplished by periodically modulating the interspecies interactions $\kappa \sin(\omega t) \hat{S}_z \hat{J}_z$ and flashing on the hopping between the two single particle states of one gas at time $\omega t= 0$ for a duration of $\tau \ll 1$, then flashing on the hopping of the other gas at time $\omega t=\pi$ for the same duration.  Altogether the driving results in the 'stirring' of the FSL.  The Floquet operator can again be written in terms of an effective Hamiltonian (Appendix \ref{app:approxham})

\begin{eqnarray}
\hat{H}_\mathrm{eff} \approx -  \sum_{n,m}  &D_+(m)& \vert n,m+1 \rangle \langle n,m \vert \nonumber \\
&+& D_-(m) \vert n,m-1 \rangle \langle n,m \vert \nonumber \\
&+& e^{i 2\pi \alpha m} D_+(n) \vert n+1, m \rangle \langle n, m\vert \nonumber \\
&+& e^{-i 2\pi \alpha m} D_-(n) \vert n-1, m \rangle \langle n, m\vert, \nonumber \\
\label{eq:heff}
\end{eqnarray} 
where the new magnetic flux ratio $\alpha = \kappa/\pi\omega$ is controlled by the interaction energy and the driving frequency.  The major difference between the real-space and Fock-space effective Hamiltonians is the presence of the $D_\pm(x) =\frac{1}{2} \sqrt{\left (N/2 \mp x \right ) \left (N/2 \pm x + 1\right )}$ factors.  This means that even without the magnetic field, the Fock lattice is \textit{not} translationally invariant.  The factors have the shape of a semicircle as a function of $x$, so for $N \gg 1$ and $x \ll N/2$, they are rather flat $D_\pm(x) \approx N/4$, then as $x$ increases, they transition to a parabolic shape with a hard wall edge at $x = \pm N/2$ because $D_\pm(\pm N/2) = 0$.  

The tunneling terms in the driving sequence can be implemented by pulsing lasers at the specified times to induce tunneling between the two modes of each gas.  We assume that the interspecies interactions take place via s-wave scattering, so the periodic interaction energy is controlled by alternating between positive and negative scattering lengths corresponding to repulsive and attractive interactions, respectively.  This is achieved through Feshbach resonance \cite{chin10} between the two species and by applying an ac-magnetic field.  In the region of a Feshbach resonance, the s-wave scattering length takes the form

\begin{equation}
a_s(t) = a_0 \left (1- \frac{\Delta}{B(t) - B_\infty} \right )
\end{equation}
where $a_s$ and $a_0$ are the modified and background s-wave scattering lengths, respectively, $\Delta$ is the width of the resonance region, $B_\infty$ is the magnetic field strength of the resonance point and $B(t)$ is the ac-magnetic field strength.  If the ac-magnetic field is $B(t) = (B_\infty + \Delta) + \delta B \sin(\omega t)$ and $\delta B \ll \Delta$, then $a_s(t) \approx \frac{a_0 \delta B}{\Delta} \sin(\omega t)$.  Time varying interaction energies were achieved experimentally not long after the creation of the first BECs to study the effects of ramping through a Feshbach resonance point \cite{abeelen99,thalhammer06,cornish00}.  Oscillating interaction energies have been used to study the association of ultracold atoms of the same species \cite{thompson05} and of different species \cite{thalhammer08,tanzi18} and also to control the excitations in a BEC \cite{pollack10}.  Experiments have also shown that periodically driving onsite interactions of an optical lattice can lead to complete suppression of the tunneling \cite{meinert16}.  Furthermore, paired time-dependent interactions and tunneling have been proposed to generate density dependent momenta of ultracold atoms in an optical lattice \cite{greschner14}.  To avoid unwanted excitations, the strength of the driving should be adiabatically ramped up.  Heating effects from enhanced inelastic collisions around the resonance point \cite{inouye98} are reduced naturally because we shift the driving away from it by $\Delta$ with a small amplitude of oscillation.  It has also been shown that heating effects can be further reduced with stronger confinement of the trapping potential \cite{bilitewski15}.

\section{\label{Sec:Results}Results}

\subsubsection{Quasispectrum}
\begin{figure}[t]
\centering
\includegraphics[width=\columnwidth]{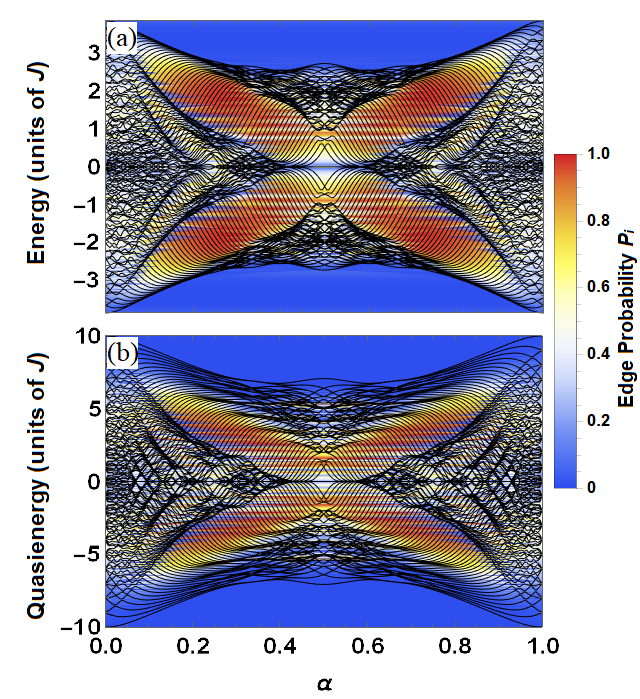}
\caption{The energy/quasienergy is plotted as a function of the magnetic flux ratio $\alpha$ and the background shows a density plot of the probability for a state of a given energy to occupy the edge of the lattice.  \textbf{(a)} The energy is calculated from the HH Hamiltonian in Eq.\ \eqref{eq:harper} with a lattice size of $11 \times 11$.  \textbf{(b)} The quasienergy is calculated from the Floquet operator in Eq.\ \eqref{eq:U2} for two gases containing 10 particles each, so the Hilbert spaces of the real- and Fock-spaces are equal.  The pulse time is $\tau = 0.01$.}
\label{fig:HB}
\end{figure}

In this section, we explore the hallmarks of magnetic fields in 2D lattices starting with the comparison of the spectra of the $\hat{H}_\mathrm{HH}$ Hamiltonian and the Floquet operator in Eq.\ \eqref{eq:U2}.  In the HH model, the magnetic field breaks translational symmetry, however, if the flux ratio is a rational number, $\alpha = p/q$ where $p,q \in \mathbb{Z}$, then the translational symmetry is restored if the unit cell increases by a factor of $q$.  The increase in size of the unit cell in real-space results in the reciprocal space being $q$ times smaller, so a single band will turn into $q$ bands.  The resulting pattern of the spectrum as a function of $\alpha$ is shown in Fig.\ \ref{fig:HB}(a) and is called Hofstadter's butterfly due to its shape.  Usually the spectrum is depicted as having a fractal structure and being without states flowing through the gaps, however, for better visualization we have used a small lattice size of $11 \times 11$ which smooths out the fractals and applied an edge in the numerics which produces edge states in the gaps.  The latter point is made clearer with the density plot in the background of the panel which is a plot of the probability for an energy eigenstate to occupy the edge of the lattice

\begin{equation}
P_i = \sum_{n,m \in \mathrm{edge}} \left | \langle n, m \vert \psi_i \rangle \right |^2 
\end{equation}
where $\vert \psi_i \rangle$ is the $i^\mathrm{th}$ energy eigenstate of $\hat{H}_\mathrm{HH}$.  The image shows that the edge states do indeed occur in the gaps with the largest edge probability being $P_\mathrm{max} \approx 0.9$, so the states are quite localized.

The set of eigenvalues of Eq.\ \eqref{eq:heff}, are called quasienergies because they come from the Floquet operator in Eq.\ \eqref{eq:U2} whose set of eigenvalues is $\{\lambda_i = \mathrm{e}^{-\mathrm{i} \epsilon_i \tau}\}$ and so lose their uniqueness when $\epsilon_i \tau > 2 \pi$.  However, when $\tau$ is small enough, quasienergies play a similar role in periodically driven systems as energies do in time independent Hamiltonians.  We plot $\frac{\mathrm{i}}{\tau} \mathrm{ln}(\{\lambda_i\})$ as a function of $\alpha$ in Fig.\ \ref{fig:HB}(b) which shows all of the qualitative features of Hofstadter's butterfly have been replicated and the lack of translational invariance of the bare FSL only has an effect on the finer details of the quasispectrum.  The edge states are again found in the gaps with a maximum probability of $P_\mathrm{max} \approx 0.8$.  The difference in the ranges along the $y$ axis between both panels comes from the $D_\pm(x)$ factors whose ranges are proportional to $N$.  

\begin{figure}[t]
\centering
\includegraphics[width=\columnwidth]{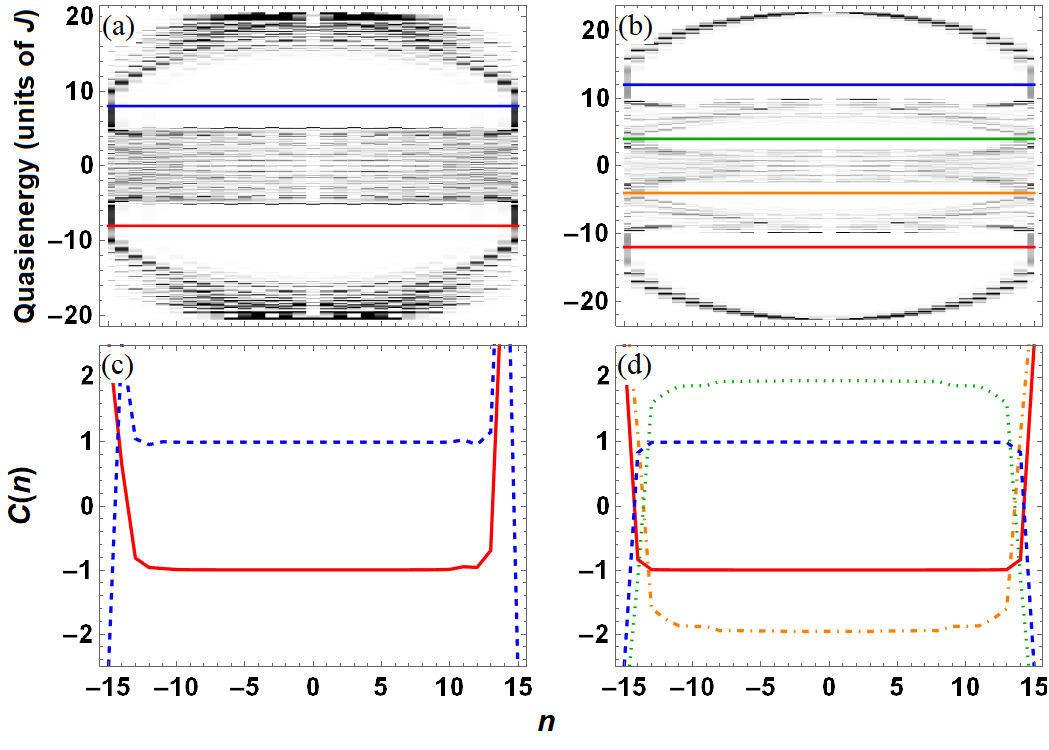}
\caption{Density plots of the probability for a site on the FSL to have a given quasienergy and the LCM for different fictitious Fermi energies ($E_F$).  \textbf{(a)} and \textbf{(b)} show $P_{\epsilon_i}(n)$ for $\alpha=1/3$ and $\alpha =1/5$, respectively, where the colored horizontal lines represent values of $E_F$.  \textbf{(c)} and  \textbf{(d)} show the LCM, $C(n)$, for the values of $E_F$ in \textbf{(a)} and \textbf{(b)}, respectively.  The density plots show bulk bands, gaps and edge states and the number of them corresponds to the value $\alpha$.  The LCM agrees with the Chern number of the HH model with $C = \pm 1$ for $\alpha = 1/3$ and $C = \pm 1, \pm 2$ for $\alpha = 1/5$ for each band.  The number of particles in each gas is $N = 30$ and the pulse time is $\tau = 0.01$.}
\label{fig:Mag}
\end{figure}

\subsubsection{Local Chern Marker}

We now turn our attention to the topological features of the model.  For 2D lattices, the Chern number is a quantity which can help identify topological properties (trivial or non-trivial).  The Chern number is usually calculated by integrating the Berry curvature over the first Brillouin zone for all occupied bands.  This calculation requires the quasimomentum to be a well defined quantity which in turn requires translational invariance.  For non-homogeneous systems, a real space version of the Chern number called the local Chern marker (LCM) has been introduced \cite{bianco11} which defines it locally in a unit cell of the lattice

\begin{equation}
C = \frac{4\pi}{A_c} \mathrm{Im} \left ( \mathrm{tr}_\mathrm{cell} \{\hat{P}\hat{x}\hat{P}\hat{y} \} \right ) 
\label{eq:LCM1}
\end{equation}
where $A_c$ is the area of the unit cell and $\hat{P}$ is the projection operator onto the states of the occupied bands.  It is expected that for large enough system sizes the average of the LCM of the bulk bands matches the quasimomentum version of the Chern number.  We neglect the trace over the unit cell because we will average the LCM over states of the bulk later on, so $A_c = 1$ and the LCM becomes completely localized

\begin{equation}
C(n,m) = - 4 \pi \mathrm{i} \langle n, m \vert \hat{P} \hat{S}_z \hat{P} \hat{J}_z \vert n, m \rangle \, .  
\end{equation}
What remains is to determine which states $\hat{P}$ projects onto and for that we need to know if the model contains any band-like structures.  To this end, we calculate the probability for a position on the lattice to have a given quasienergy $\epsilon_i$ (calculated from the Floquet operator in Eq.\ \eqref{eq:U2}), $P_{\epsilon_i}(n,m) = \vert \langle n, m \vert \epsilon_i \rangle \vert^2$, and plot an $m = 0$ slice of it, $P_{\epsilon_i}(n) = P_{\epsilon_i}(n,m=0)$, for two values of the flux ratio $\alpha$ in Fig.\ \ref{fig:Mag} (a) and (b).  Panel (a) shows three distinct band-like structures for $\alpha = 1/3$ and panel (b) shows five for $\alpha = 1/5$.  Although these are not energy bands in the strictest sense, they still follow the qualitative feature of the HH model that there are $q$ bands when $\alpha  = 1/q$.  Each band is separated by a gap with the exception of the edges of the lattice at $n = \pm 15$ ($N = 30$ for both gases) where the bands are connected agreeing with the existence of edge states in the gaps in Fig.\ \ref{fig:HB}.

In normal real-space lattices there is a Fermi surface with an energy $E_F$ where states with energy below $E_F$ are occupied by electrons.  It is these states that belong to the projection operator in Eq.\ \eqref{eq:LCM1}.  Such a Fermi surface does not exist for the FSL, however, we can imagine a fictitious one existing in the gaps between bands to help facilitate the calculation of the LCM.  Different Fermi energies are shown as horizontal colored lines in panels (a) and (b) of Fig.\ \ref{fig:Mag} and are placed at an arbitrary position within the gaps.  Therefore, we will take $\hat{P}$ as the projection operator of states with energies below these lines.  Panels (c) and (d) show $C(n) = C(n,0)$ for the Fermi energies in (a) and (b), respectively.  One can see that the LCM varies significantly over the edges, but settles down to a constant value farther inside the bulk.  We take the average of the LCM in the region $\vert n \vert, \vert m \vert \leq 8$, so that it is taken over a region comfortably inside the bulk.  The average LCMs for increasing Fermi energy are: $C_\mathrm{ave} = $ -0.996, 0.996 for $\alpha = 1/3$ and $C_\mathrm{ave} = $ -0.998, -1.934, 1.934, 0.998 for $\alpha = 1/5$.  These values agree well with the Harper model Chern numbers of $\pm 1$ for $\alpha = 1/3$ and $\pm 1, \pm 2$ for $\alpha = 1/5$ and satisfy the condition that the sum of the Chern numbers for all of the bands is equal to zero.

\subsubsection{Intraspecies Interactions}

\begin{figure}[t]
\centering
\includegraphics[width=\columnwidth]{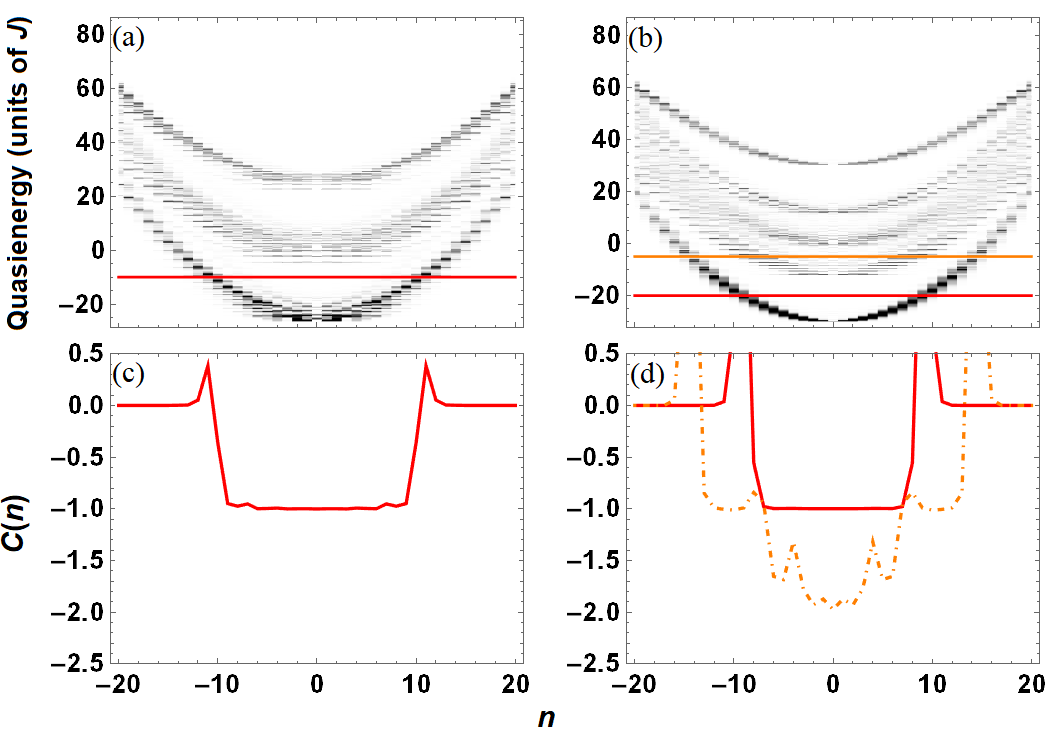}
\caption{The same images as in Fig.\ \ref{fig:Mag} except with intraspecies interactions $\mu = 0.02$ and the number of particles in each gas is now $N = 40$ instead of $N = 30$ to make the gaps more visible.  When $\mu \neq 0$ the intraspecies interactions act as a harmonic trap on the FSL causing the parabolic shape of the bands.  The LCM shows that for weak interactions the bands retain their topological features with values matching the Chern numbers of the HH model in the bulk. }
\label{fig:Mag3}
\end{figure}

Intraspecies interactions are introduced with the term $U\left ( \hat{S}_z^2 + \hat{J}_z^2 \right )$.  For simplicity we take the interaction energy $U$ to be the same for both gases and it is controlled by the s-wave scattering length of the particles.  Only repulsive interactions $U > 0$ will be considered, so their inclusion results in a harmonic trap in the FSL.  The evolution operator in Eq.\ \eqref{eq:U2} gains the factor  $\mathrm{e}^{- \mathrm{i} \frac{\mu}{N} \left (\hat{S}_z^2 + \hat{J}_z^2 \right )}$ at the first and third positions (left to right) where $\mu = UN\pi/\omega$.  Harmonic traps have been studied for real-space lattices \cite{buchhold12} because they are needed as an overall trapping potential in experimental setups involving ultracold atomic gases.  It was found that harmonic confinement creates a soft-wall boundary for the lattice which makes edge states harder to identify due to smearing.  However, bulk bands were still visible, as well as gaps between them for large enough lattices and weak enough confinements, so topological features still persisted.  We plot $P_{\epsilon_i}(n)$ with $\mu \neq 0$ for the two values of $\alpha$ in Fig.\ \ref{fig:Mag3} (a) and (b) and find that the soft wall of the harmonic trap does indeed smear out the bands near the edges, however, they still connect bands together due to the hard wall at the very edge.  The interactions also decrease the gap size in the longitudinal direction.  This suggests that there is a critical interaction strength for a finite sized system where the bands get so smeared out that the gaps disappear.  Even so, for weak enough interactions the bands and gaps are still distinguishable which is what we see for $\mu = 0.02$.  

In Fig.\ \ref{fig:Mag3} (c) and (d) we once again plot the LCM, $C(n)$, for the fictitious Fermi energies in (a) and (b), respectively, however, we only include the lower energy LCMs since we have established that the upper ones are the same, but with a flipped sign.  With interactions we find that the average LCM values are $C_\mathrm{ave} = -0.987$ for $\alpha = 1/3$ and $C_\mathrm{ave} = -0.996, -1.918$ for $\alpha = 1/5$.  These values again agree with the HH model Chern numbers signaling that although the interactions change the shape of the bands, their topological properties remain intact.  Away from the center of the lattice there are some qualitative differences compared to the case with no interactions.  In (c) the low Fermi energy LCM (red, smooth) quickly goes to zero because, due to the curvature of the bands, $E_F$ enters a region where there are no states as seen in (a).  This crossing occurs at $n = \pm 12$.  The higher Fermi energy LCM in (d) (orange, dot-dashed) varies significantly away from the center because $E_F$ crosses the second bulk band and enters the lower gap at $n = \pm 10$, so the LCM briefly picks up the lower band LCM value of $-1$.  Near the edges of the lattice, the LCM goes to zero again because there are no more states below $E_F$ in this area.

\subsubsection{Transport Properties}

In condensed matter systems the Chern number is usually measured through the Hall conductance \cite{klitzing80}, however, it is difficult to use the same technique for neutral particles in an ultracold atomic gas.  One proposed method uses hybrid time-of-flight (HTOF) images \cite{wang13} which involves \textit{in situ} measurements of a gas cloud's density in one direction of the lattice and measurements of the free expansion of the gas in the other direction.  The HTOF images are quantified in the density of particles with a given position and transverse quasimomentum, $\rho(k_x, y)$.  This quantity takes advantage of the fact that the Chern number is related to adiabatic transport properties of the lattice and that a 2D lattice can also be thought of as a 1D charge pump where $k_x$ plays the role of the pumping parameter.  When $k_x$ is cycled over the Brillouin zone, $k_x \to k_x + 2\pi$, the particles on the lattice are displaced if there is nontrivial topology present.  The displacement is related to the Chern number by $\delta y_\mathrm{COM} = Cq$ where $\delta y_\mathrm{COM}$ is the center-of-mass displacement, $C$ is the Chern number and $q$ is the length of the magnetic unit cell in the $y$ direction when $\alpha = 1/q$.  

\begin{figure}[t]
\centering
\includegraphics[width=\columnwidth]{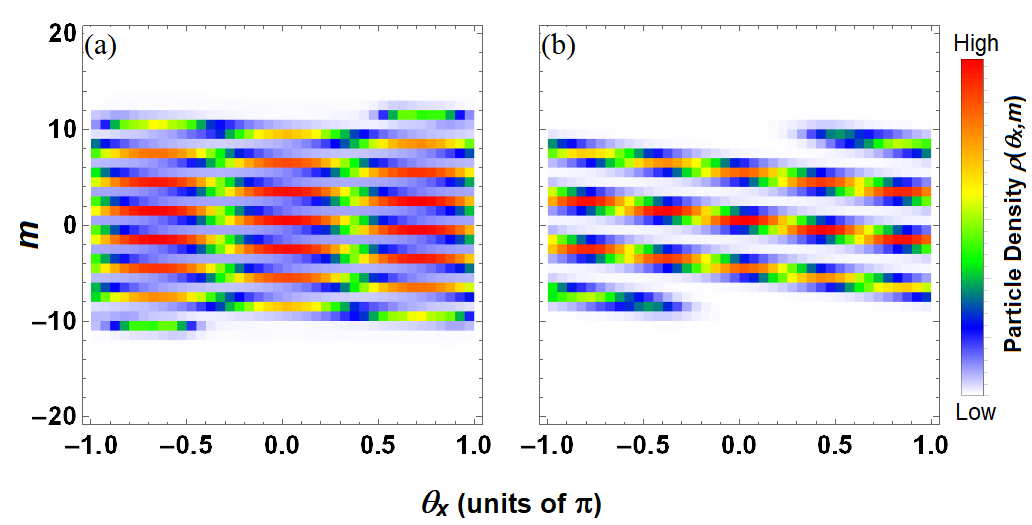}
\caption{The particle density function, $\rho(\theta_x,m)$, is plotted as a function of the relative phase of one gas and the number difference of the other gas.  Panels \textbf{(a)} and \textbf{(b)} show $\rho(\theta_x,m)$ for the lowest bands for $\alpha = 1/3$ and $\alpha = 1/5$, respectively.  A peak of the density is displaced by -3 in \textbf{(a)} and -5 in \textbf{(b)} agreeing with the charge pump result of $\delta m = Cq$ in real-space lattices.  The other parameter values match those in Fig.\ \ref{fig:Mag3}.}
\label{fig:Mag4}
\end{figure}

In the current system, the relative phases between states 1 and 2 of each BEC, $\theta_x$ and $\theta_y$, are the conjugate variables to the number differences $n$ and $m$, respectively, so they are analogous to the quasimomenta in real-space.  In the Schwinger representation, the relative phase is also the azimuthal angle on the surface of a spin-$N/2$ Bloch sphere.  The particle density takes the form

\begin{equation}
\rho(\theta_x, m) = \frac{1}{\mathcal{N}} \sum_{i = 1}^\mathcal{N} \vert \Phi_i(\theta_x, m)\vert^2
\label{eq:PD}
\end{equation}  
where the sum is over the $\mathcal{N}$ states below the Fermi energy and 

\begin{equation}
\Phi_i(\theta_x, m) = \left ( N+1 \right )^{-1/2} \sum_{n = -N/2}^{N/2} \mathrm{e}^{\mathrm{i} \theta_x n} \psi_i(n,m)
\end{equation}
is the Fourier transform of the $\mathrm{i}^\mathrm{th}$ eigenstate of the Floquet operator.  The relative phase also takes discrete values, $\theta_x = \frac{2\pi r}{N+1}$ where $r$ is an integer in the range $-N/2 \leq r \leq N/2$.  Similarly to the real-space transport, we expect there to be a displacement of $\delta m = Cq$ on the FSL when $\theta_x$ changes by $2\pi$.  

We will consider the lowest Fermi energy (red) in Fig.\ \ref{fig:Mag3} for both values of $\alpha$ which means $C = -1$.  In one cycle of $\theta_x$ the expected displacement is $\delta m = -3$ for $\alpha = 1/3$ and $\delta m = -5$ for $\alpha = 1/5$.  Figure \ref{fig:Mag4} shows $\rho(\theta_x,m)$ and we find agreement with the predicted result.  Following a peak of $\rho(\theta_x,m)$ (red) as $\theta_x$ increases by $2\pi$ results in the peak being displaced by $\delta m = -3$ for $\alpha = 1/3$ in (a) and by $\delta m = -5$ for $\alpha = 1/5$ in (b).  A helpful guide to identify the displacement in the figure is to note that $m$ is discrete, so the images are broken up into horizontal strips.  The number of strips going from one peak to another is three in (a) and five in (b).  The downward slope is attributed to the Chern number being negative, however, a Chern number of $+1$ can also be checked for the lowest band by flipping the sign of the interspecies interactions $\kappa$ in Eq.\ \eqref{eq:U2}.  The intraspecies interactions do not have an effect on the transport properties of the FSL except to prevent any displacement near the sides of the lattice where the sites are energetically inaccessible (white-space in the images).  

Like $k_x$, $\theta_x$ can also be measured from time-of-flight expansion through the interference fringes of absorption images \cite{andrews97}.  This means that in order to see the transport effects one must be able to release one gas from the trap for the $\theta_x$ measurement and keep the other gas trapped for the $n$ measurement.  One of the major difficulties in performing these measurements is the requirement that the gases be separate, however, in general, they will end up mixing.  One way to aid in the measurements is if the two modes of each gas are spatially separated like the two wells of a BJJ where the tunneling is through a central potential barrier.  On the FSL, quadrants two and four represent immiscible states of the BJJ, so measurements taken in these regions will have some separation between the gases. 

\begin{figure}[t]
\centering
\includegraphics[width=\columnwidth]{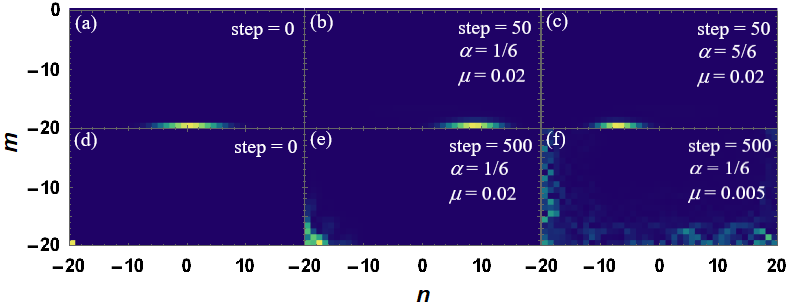}
\caption{Initial and final probability distributions in the FSL for different parameter values. \textbf{(a)} An initial Gaussian distribution (spin coherent state) along the horizontal centered at $(n,m) = (0,-N/2)$.  \textbf{(b)} After 50 steps of the Floquet operator in Eq.\ \eqref{eq:U2} the Gaussian has moved along the edge to the right while maintaining its shape for $\alpha = 1/6$ and $\mu = 0.02$.  \textbf{(c)} The same as \textbf{(b)} except the flux ratio has been changed to $\alpha = 5/6$ which reverses the chirality of the Gaussian, so it has moved along the edge to the left after 50 steps.  \textbf{(d)} An initial Fock distribution located in the bottom left corner at $(n,m) = (-N/2,-N/2)$. \textbf{(e)} After 500 steps of the Floquet operator the distribution remains in its initial vicinity for $\alpha = 1/6$ and $\mu = 0.02$ which is a qualitative sign of MQST.  \textbf{(f)} The interaction energy is reduced to $\mu = 0.005$ ceasing MQST and after 500 steps the distribution has spread along the left and bottom edges.  The images were generated using $N = 40$ for both gases and with a pulse time of $\tau = 0.01$.}
\label{fig:chiral}
\end{figure}

Staying with the BJJ system, the edge state chirality can be measured by preparing the initial state as a spin coherent state in the $x$-direction and a Fock state at $m = -N/2$ in the $y$-direction.  This results in a Gaussian state in the $x$-direction centered at $(n,m) = (0,-N/2)$ on the lattice.  Such a state can be prepared by having relatively strong tunneling between the two wells of the $x$-direction gas and loading all of the particles of the $y$-direction gas into one well (well 2 in this case) before the driving starts.  This initial state is shown at the bottom edge of panel (a) in Fig.\ \ref{fig:chiral}.  Panel (b) shows the wavefunction has moved to the right along the edge after 50 steps of $\hat{U}$ for parameter values $\alpha = 1/6$ and $\mu = 0.02$.  The chirality can be flipped by flipping the sign of $\kappa$ ($\alpha$), as previously mentioned.  Equivalently, due to the symmetry of the quasispectrum in Fig.\ \ref{fig:HB}, the chirality can be flipped by setting $\alpha = 5/6$ which is shown in panel (c) where the wavefunction has now moved to the left.  The bottom right and left corners of the Fock lattice represent complete separation and complete overlap of the two gases in the BJJ, so the chirality can be inferred by the miscibility of the final state.  Tunable miscibility experiments have been conducted by controlling the Feshbach resonances of both the interspecies and intraspecies interactions of quantum gases \cite{wang16,lee18}.  

A distinguishing feature of the single species BJJ is a dynamical process called macroscopic quantum self-trapping (MQST).  This process comes from the intraspecies interactions and leads to $n$ having a nonzero long time average.  In extreme cases when the interactions are large enough, the number difference does not evolve much at all.  This is a rather counterintuitive result because one would think that for larger $\mu$ (more repulsive) the gas would tend toward a state that is maximally spread between the two wells, not prefer one well over the other.  Clarification of MQST can be gained by considering the classical rigid pendulum.  If given a large enough kinetic energy, the pendulum will swing over the top and make full loops as it evolves.  For a very large kinetic energy, the pendulum swings around rapidly maintaining a near constant angular momentum.  This happens because there is no angular displacement state that makes the gravitational potential energy comparable to the kinetic energy, so the kinetic energy and therefore the angular momentum are locked.  In the BJJ, the intraspecies interaction energy and the boson number difference $n$ ($m$) are analogous to the kinetic energy and the angular momentum of the pendulum, respectively.  Therefore, small oscillations of the BEC moving back and forth between the two wells is similar to small oscillations of the pendulum.  Likewise, the pendulum having enough kinetic energy to make full loops to maintain a nonzero angular momentum is the same as the BEC having enough intraspecies interaction energy to maintain a nonzero boson number difference and some of the BEC ends up trapped in one well over the other. 

\begin{figure}[t]
\centering
\includegraphics[width=\columnwidth]{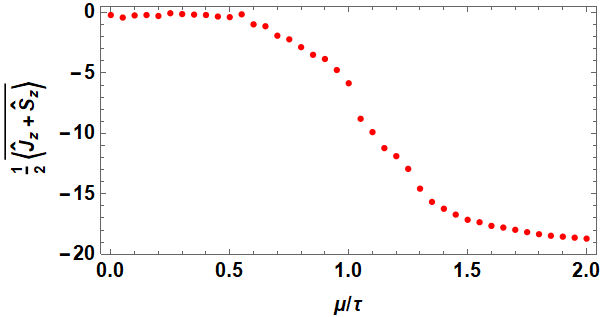}
\caption{The long-time average of $\frac{1}{2} \langle \hat{S}_z+ \hat{J}_z\rangle (t)$ (denoted by the overline in the image) for different values of the ratio between the intraspecies interaction energy and the pulse time, $\mu/\tau$.  The brackets $\langle ... \rangle$ denote the expectation value with respect to the Fock state $\vert -N/2, - N/2 \rangle$ shown in Fig.\ \ref{fig:chiral} (d).  For small values of the ratio the state spreads along the edges of the lattice and the long-time average is zero.  Around the critical ratio of $\left ( \mu/\tau \right )_\mathrm{c} = 1$ MQST begins to take effect stopping the wavefunction from spreading, so the long-time average is close to its initial value of $-N/2$.  The parameter values are $N = 40$ for both gases, $\tau = 0.01$ and the time average was taken over 10000 steps of $\hat{U}$. }
\label{fig:LTA}
\end{figure}

To show MQST in the current system we change the initial state to the Fock state $\vert -N/2, -N/2 \rangle$ which corresponds to both gases being loaded into the same well before the driving starts.  This state can be seen in the bottom left corner of Fig.\ \ref{fig:chiral} (d).  Panel (e) shows the wavefunction after 500 steps of $\hat{U}$ for parameter values of $\alpha = 1/6$ and $\mu = 0.02$.  We can see that MQST has prevented the wavefunction from leaving the corner (confirmed for all intermediate steps) resulting in a localized state on the FSL.  By reducing $\mu$ we reduce the amount of energy stored in the intraspcecies interactions and the wavefunction will be able to explore more of the lattice.  This is what we see in panel (f) where we keep the same $\alpha$, but now $\mu = 0.005$.  The localization can also be explained through the examination of the high energy eigenstates of $\hat{U}$ as was done for the HH model in a harmonic trap \cite{kolovsky14}.  It was found that as $\mu$ (strength of the harmonic trap) increases, the higher energy states form a near four-fold degenerate group and the initial state $\vert -N/2,-N/2\rangle$ can be approximately constructed from a superposition of these states.  The result is that the initial state is close to being an eigenstate of $\hat{U}$, so it does not evolve much.

To find the critical value of $\mu$ where the onset of MQST occurs we derive the meanfield Hamiltonian of $\hat{H}_\mathrm{eff}$ with interactions giving  (Appendix \ref{app:mfham})

\begin{align}
H_\mathrm{MF} =  \frac{\mu}{\tau} & \left (x^2 + y^2 \right ) - \left [ \sqrt{1-y^2} \cos(\theta_y) \right . \nonumber \\
& + \left . \sqrt{1-x^2} \cos ( \theta_x + \pi\alpha Ny) \right ] 
\label{eq:MF}
\end{align}
where $x = 2n/N$ and $y = 2m/N$.  If we imagine both $x$ and $y$ as angular momentum variables ($\theta_x$ and $\theta_y$ as angular displacement variables), then $H_\mathrm{MF}$ resembles the Hamiltonian of a system of coupled pendula, albeit a strange one due to the angular momentum dependent potential energy.  Nevertheless, we can use the pendulum analogy discussed earlier to say that MQST should take effect when the coupled pendula in Eq.\ \eqref{eq:MF} have enough energy to make a full loop, or in other words, when the total energy is greater than the maximum potential energy.  The initial state is located at $(n,m) = (-N/2,-N/2)$, or $(x,y)=(-1,-1)$, so its energy is $2\mu/\tau$.  The maximum potential energy occurs at $(x,y,\theta_x,\theta_y) = (0,0,\pi,\pi)$ and is equal to 2, so MQST happens when $2\mu/\tau \geq 2$ giving a critical value of

\begin{equation}
\left (\frac{\mu}{\tau} \right )_\mathrm{c} = 1 \, .
\end{equation} 
We stress that the critical value depends on the initial state as different initial states have different total energies.  In fact, the dependence of the critical value on $\mu$ and $\alpha$ can be quite complicated in some systems as seen in the periodically driven BEC in a quadruple well \cite{wang21}.  Figure \ref{fig:LTA} shows $\frac{1}{2}\overline{ \langle \hat{S}_z + \hat{J}_z \rangle (t)}$, which is the the long-time average (denoted by the line) of the expectation value of the sum of the number difference operators, for different values of the ratio $\mu/\tau$.  The expectation is taken with respect to the Fock state $\vert -N/2, -N/2 \rangle$ shown in the bottom left corner of Fig.\ \ref{fig:chiral} (d).  The image shows that for small values of $\mu/\tau$ the long-time average is zero because the state spreads along the edges of the entire lattice as it evolves.  Around the critical ratio of unity the long-time average jumps signifying the onset of MQST and asymptotically approaches $-N/2$.  This analysis has highlighted the key difference between the Fock-space and real-space lattices which is that the sites on the FSL are fundamentally many-body states and the real space sites are not.  Thus, many-body phenomena like MQST can appear on the FSL and in this case disrupt edge states.
 
\section{\label{Sec:Conc}Conclusion}

In this work we showed that two interacting bosonic quantum gases can be periodically driven to simulate a particle on a 2D lattice in the presence of a magnetic field.  The quasispectrum of the driving sequence Floquet operator reproduces the celebrated Hofstadter's butterfly pattern found in the HH model.  Through the calculation of the LCM we found that the driven gases retain the topological features of the HH model with the Chern numbers agreeing between the two systems.  A method to measure the topological properties of the driven gases is proposed involving a combination of measuring the relative phase of one gas and the number difference of the other gas.  This method is akin to hybrid time-of-flight images and measures the transport properties of the lattice associated with the Chern number.  Finally, we showed that when intraspecies interactions are included, the nonlinear effect of MQST can hinder the movement of edge states.

The implementation of the driving sequence raises some interesting possibilities for the types of synthetic lattices one can build.  For instance, instead of coupling two many-particle two-mode gases together, only the addition of a \textit{single} two-mode particle to one $N$ particle gas is enough for a second dimension.  The FSL then takes the shape of a ladder with the rungs made of the two-mode single particle and the legs made of the two-mode $N$ particle gas.  Also, a single three-mode gas has a hexagonal shaped Hilbert space with a triangular hard wall, so this may open up possibilities to explore other important systems like the Haldane model.  Further study of simulating lattices with ultracold atomic gases will provide opportunities to explore the relation between seemingly disparate phenomena in the condensed matter, optical and atomic fields. 

\acknowledgments  \textbf{Acknowledgements} J. Mumford thanks J. Larson, W. Kirkby, and D. H. J. O'Dell for helpful comments in the writing of this manuscript.


%

%

\appendix

\section{\label{app:approxham} Derivation of $\hat{H}_\mathrm{eff}$}

To derive Eq.\ \eqref{eq:heff} we start with the unitary Floquet operator in Eq.\ \eqref{eq:U2} and using the fact that $\mathrm{e}^{\mathrm{i} \phi \hat{J}_z} f\left (\hat{J}_x \right ) \mathrm{e}^{-\mathrm{i} \phi \hat{J}_z} = f\left (\hat{J}_x \cos(\phi) - \hat{J}_y \sin(\phi) \right )$ where $f(x)$ is a general function of $x$ gives

\begin{equation}
\hat{U} = \mathrm{e}^{\mathrm{i}\tau \left [ \hat{J}_x \cos\left (2\pi \alpha \hat{S}_z \right ) - \hat{J}_y \sin \left ( 2\pi \alpha \hat{S}_z \right )\right ]} \mathrm{e}^{\mathrm{i}\tau\hat{S}_x}
\label{eq:UA}
\end{equation}
where the substitution $\alpha = \kappa/\omega \pi$ has been made.  Using the Baker-Campbell-Hausdorff formula

\begin{equation}
\mathrm{e}^{\mathrm{i} \tau \hat{A}} \mathrm{e}^{\mathrm{i} \tau \hat{B}} = \mathrm{e}^{\mathrm{i} \tau \left (\hat{A}+ \hat{B} \right ) + \frac{\left (\mathrm{i}\tau \right )^2}{2} \left [ \hat{A}, \hat{B} \right ] + ... }
\end{equation}
and taking $\tau \ll 1$, so that terms of $\mathcal{O}\left ( \tau^2 \right )$ and higher can be neglected gives

\begin{equation}
\hat{U} \approx \mathrm{e}^{\mathrm{i}\tau \left [ \hat{J}_x \cos\left (2\pi \alpha \hat{S}_z \right ) - \hat{J}_y \sin \left ( 2\pi \alpha \hat{S}_z \right ) + \hat{S}_x \right ]} \, .
\end{equation}
Writing the Floquet operator as $\hat{U} = \mathrm{e}^{-\mathrm{i} \hat{H}_\mathrm{eff} \tau}$, we identify the effective Hamiltonian as 

\begin{equation}
\hat{H}_\mathrm{eff} = -\left [\hat{J}_x \cos\left (2\pi \alpha \hat{S}_z \right ) - \hat{J}_y \sin \left ( 2\pi \alpha \hat{S}_z \right ) + \hat{S}_x \right ] \, .
\end{equation}
In terms of the spin raising and lowering operators, the Hamiltonian becomes

\begin{equation}
\hat{H}_\mathrm{eff} = -\frac{1}{2} \left [ \hat{J}_+ \mathrm{e}^{\mathrm{i} 2\pi \alpha \hat{S}_z } + \hat{J}_- \mathrm{e}^{-\mathrm{i} 2\pi \alpha \hat{S}_z } + \hat{S}_+ +\hat{S}_- \right ] 
\end{equation}
and expressing the spin operators explicitly in terms of the $\hat{J}_z$ ($\hat{S}_z$) eigenstates produces the effective Hamiltonian in Eq.\ \eqref{eq:heff}.

\section{\label{app:mfham} Derivation of $H_\mathrm{MF}$}

We start the derivation of $H_\mathrm{MF}$ by adding intraspecies interactions to the effective Hamiltonian.  With intraspecies interactions the unitary operator in Eq.\ \eqref{eq:UA} becomes 

\begin{eqnarray}
\hat{U} = &\mathrm{e}^{- \mathrm{i} \frac{\mu}{N} \left (\hat{S}_z^2 + \hat{J}_z^2 \right )}\mathrm{e}^{\mathrm{i}\tau \left [ \hat{J}_x \cos\left (2\pi \alpha \hat{S}_z \right ) - \hat{J}_y \sin \left ( 2\pi \alpha \hat{S}_z \right )\right ]} \nonumber \\
& \times \mathrm{e}^{- \mathrm{i} \frac{\mu}{N} \left (\hat{S}_z^2 + \hat{J}_z^2 \right )} \mathrm{e}^{\mathrm{i}\tau\hat{S}_x} \, .
\end{eqnarray}
In this paper we are using small values of $\mu$ of $\mathcal{O}(\tau)$, so we can neglect terms of $\mathcal{O}(\mu \tau)$ in the Baker-Campbell-Hausdorff formula and simply combine all of the exponentials into one

\begin{equation}
\hat{U} = \mathrm{e}^{\mathrm{i} \tau \left \{-\frac{2\mu}{N\tau} \left (\hat{S}_z^2 + \hat{J}_z^2 \right ) + \hat{J}_x \cos \left (2\pi\alpha \hat{S}_z\right )-\hat{J}_y \sin\left (2\pi\alpha\hat{S}_z \right )+ \hat{S}_x \right \}} \, ,
\end{equation}
so the effective Hamiltonian with interactions is

\begin{eqnarray}
\hat{H}_\mathrm{eff} = \frac{2\mu}{N\tau} &\left (\hat{S}_z^2 + \hat{J}_z^2 \right ) - \hat{J}_x \cos \left (2\pi\alpha \hat{S}_z\right ) \nonumber \\
& +\hat{J}_y \sin\left (2\pi\alpha\hat{S}_z \right )- \hat{S}_x \, .
\end{eqnarray}
The mean-field effective Hamiltonian is derived by replacing the creation (annihilation) operators with complex numbers

\begin{equation}
\hat{a}_{1,2} \to \sqrt{N_{1,2}} \mathrm{e}^{\mathrm{i} \theta_{1,2}}, \hspace{15pt} \hat{b}_{1,2} \to \sqrt{M_{1,2}} \mathrm{e}^{\mathrm{i} \phi_{1,2}}
\end{equation}
which transforms the spin operators into

\begin{eqnarray}
&\hat{J}_z \to \frac{1}{2} \left (N_1 - N_2 \right ) = n, \, \hat{S}_z \to \frac{1}{2} \left (M_1 - M_2 \right ) = m \nonumber \\
&\hat{J}_x \to \sqrt{N^2/4 - n^2} \cos\left (\theta_x \right ), \, \hat{S}_x \to \sqrt{N^2/4 - m^2} \cos\left (\theta_y \right ) \nonumber \\
&\hat{J}_y \to \sqrt{N^2/4 - n^2} \sin\left (\theta_x \right ), \, \hat{S}_y \to \sqrt{N^2/4 - m^2} \sin\left (\theta_y \right )\nonumber \\
\end{eqnarray}
where $\theta_x = \theta_1 - \theta_2$ and $\theta_y = \phi_1 - \phi_2$.  We take the mean-field Hamiltonian to be $H_\mathrm{MF} = \lim_{N\to\infty} \hat{H}_\mathrm{eff}/(N/2)$, so it becomes

\begin{eqnarray}
H_\mathrm{MF} = &\frac{\mu}{\tau} \left (x^2 + y^2 \right ) - \sqrt{1-x^2} \cos\left (\theta_x + \pi\alpha Ny \right ) \nonumber \\
&- \sqrt{1-y^2} \cos\left ( \theta_y \right )
\end{eqnarray}
where $x = 2n/N$ and $y = 2m/N$ and we have used the trigonometric identity $\cos\left (a+b \right ) = \cos (a)\cos(b) - \sin(a) \sin(b)$.


\begin{thebibliography}{8}

\bibitem{jaksch05}{D. Jaksch and P. Zoller, The cold atom Hubbard toolbox, Ann. Phys. \textbf{315}, 52 (2005); I. Bloch, J. Dalibard, and W. Zwerger, Many-body physics with ultracold gases, Rev. Mod. Phys. \textbf{80}, 885 (2008).}

\bibitem{dalibard11}{J. Dalibard, F. Gerbier, G. Juzeli$\bar{\mathrm{u}}$nas, and P \"{O}hberg, Colloquium: Artificial Gauge Potentials for Neutral Atoms, Rev. Mod. Phys. \textbf{83}, 1523 (2011).}

\bibitem{goldman14a}{N. Goldman, G. Juzeli$\bar{\mathrm{u}}$nas, P. \"{O}hberg, and I. B. Spielman, Light-induced gauge fields for ultracold atoms, Rep. Prog. Phys. \textbf{77}, 126401 (2014).}

\bibitem{goldman14b}{N. Goldman and J. Dalibard, Periodically driven quantum systems: effective Hamiltonians and engineered gauge fields, Phys. Rev. x \textbf{4} 031027 (2014).}

\bibitem{eckardt15}{A. Eckardt and E. Anisimovas, High-frequency approximation for periodically driven quantum systems from a Floquet space perspective, New J. Phys. \textbf{17}, 093039 (2015).}

\bibitem{burkov15}{M. Burkov, L. D'Alessio, and A. Polkovnikov, Universal high-frequency behavior of periodically driven systems: from dynamical stabilization to Floquet engineering, Adv. Phys. \textbf{64}, 139-226 (2015).}

\bibitem{creffield16}{C. E. Creffield, G. Pieplow, F. Sols, and N. Goldman, Realization of uniform synthetic magnetic fields by periodically shaking an optical square lattice, New J. Phys. \textbf{18}, 093013 (2016).}

\bibitem{madison00}{K. W. Madison, F. Chevy, W. Wohlleben, and J. Dalibard, Vortex formation in a stirred Bose-Einstein condensate, Phys. Rev. Lett. \textbf{84}, 806 (2000); J. Abo-Shaeer, C. Raman, J. Vogels, and W. Ketterle, Observation of vortex lattices in Bose-Einstein condensates, Science \textbf{292}, 476-479 (2001).}

\bibitem{schweikhard04}{V. Schweikhard, I Coddington, P. Engels, V. P. Mogendorff, and E. A. Cornell, Rapidly rotating Bose-Einstein condensates in and near the lowest Landau level, Phys. Rev. Lett. \textbf{92}, 040404 (2004).}

\bibitem{aidelsburger11}{M. Aidelsburger, M. Atala, S. Nascimb\'{e}ne, S. Trotzky, Y.-A. Chen, and I. Bloch, Experimental realization of strong effective magnetic fields in an optical lattice, Phys. Rev. Lett. \textbf{107}, 255301 (2011).}

\bibitem{struck11}{J. Struck, C. \"{O}lschl\"{a}ger, R. Le Targat, P. Soltan-Panahi, A. Eckardt, M. Lewenstein, P. Windpassinger, and K. Sengstock, Quantum simulation of frustrated classical magnetism in triangular optical lattices, Science \textbf{333}, 996-999 (2011).}

\bibitem{aidelsburger13}{M. Aidelsburger, M. Atala, M. Lohse, J. T. Berreiro, B. Paredes, and I Bloch, Realization of the Hofstadter Hamiltonian with ultracold atoms in optical lattices, Phys. Rev. Lett. \textbf{111}, 185301 (2013).}

\bibitem{miyake13}{H. Miyake, G. A. Siviloglou, C. J. Kennedy, W. C. Burton, and W. Ketterle, Realizing the Harper Hamiltonian with laser-assisted tuennling in optical lattics, Phys. Rev. Lett. \textbf{111}, 185302 (2013).}

\bibitem{jotzu14}{G. Jotzu, M. Messer, R. Desbuquois, M. Lebrat, T. Uehlinger, D. Greif, and T. Esslinger, Experimental realization of the topological Haldane model with ultracold fermions, Nature \textbf{515}, 237-240 (2014).}


\bibitem{boada12}{O. Boada, A. Celi, J. I. Latorre, and M. Lewenstein, quantum simulation of an extra dimension, Phys. Rev. Lett. \textbf{108}, 133001 (2012).}

\bibitem{ozawa19b}{T. Ozawa, H. M. Price, Topological quantum matter in synthetic dimensions, Nat. Rev. Phys. \textbf{1}, 349-357 (2019).}

\bibitem{celi14}{A. Celi, P. Massignan, J. Ruseckas, N. Goldman, I. B. Spielman, G. Juzeli$\bar{\mathrm{u}}$nas, and M. Lewenstein, Synthetic gauge fields in synthetic dimensions, Phys. Rev. Lett. \textbf{112}, 043001 (2014).}

\bibitem{mancini15}{M. Mancini, G. Pagano, G. Cappellini, L. Livi, M. Rider, J. Catani, C. Sias, P. Zoller, M. Inguscio, M. Dalmonte, and L. Fallani, Observation of chiral edge states with neutral fermions in synthetic Hall ribbons, Science \textbf{349}, 1510-1513 (2015).}

\bibitem{stuhl15}{B. K. Stuhl, H.-I. Lu, L. M. Aycock, D. Genkina, and I. B. Spielman, Visualizing edge states with an atomic Bose gas in the quantum Hall regime, Science \textbf{349}, 1514-1518 (2015).}

\bibitem{anisimovas16}{E. Anisimovas, R. Ra\v{c}i\={u}nas, C. Str\"{a}ter, A. Eckardt, I. B. Spielman, and G. Juzeli\={u}nas, Semisynthetic zigzag optical lattice for ultracold bosons, Phys. Rev. A \textbf{94}, 063631 (2016).}

\bibitem{an17}{F. A. An, E. J. Meier, and B. Gadway, Diffusive and arrested transport of atoms under tailored disorder, Nat. Comm. \textbf{8}, 325 (2017); F. A. An, E. J. Meier, J. Ang'ong'a, and B. Gadway, Correlated dynamics in a synthetic lattice of momentum states, Phys. Rev. Lett. \textbf{120}, 040407 (2018).}

\bibitem{meier16}{E. J. Meier, F. A. An, and B. Gadway, Observation of the topological soliton state in the Su-Schrieffer-Heeger model, Nat. Comm. \textbf{7}, 13986 (2016).}

\bibitem{xie19}{D. Xie, W. Gou, T. Xiao, B. Gadway, and B. Yan, Topological characterizations of an extended Su-Schrieffer-Heeger model, npj Quantum Inf. \textbf{5}, 55 (2019).}

\bibitem{livi16}{L. F. Livi, G. Cappellini, M. Diem, L. Franchi, C. clivati, M. Frittelli, F. Levi, D. Calonico, J. Catani, M. Inguscio, and L. Fallani, Synthetic dimensions and spin-orbit coupling with an optical clock transition, Phys. Rev. Lett. \textbf{117}, 220401 (2016).}

\bibitem{kolkowitz17}{S. Kolkowitz, S. L. Bromley, T. Bothwell, M. L. Wall, G. E. Marti, A. P. Koller, X. Zhang, A. M. Rey, and J. Ye, Spin-orbit coupled fermions in an optical lattice clock, Nature, \textbf{542}, 66-70 (2017).}

\bibitem{wall16}{M. L. Wall, A. P. Koller, S. Li, X. Zhang, N. R. Cooper, J. Ye, and A. M. Rey, Synthetic spin-orbit coupling in an optical lattice clock, Phys. Rev.Lett. \textbf{116}, 035301 (2016).}

\bibitem{price17}{H. M. Price, T. Ozawa, and N. Goldman, Synthetic dimensions for cold atoms from shaking a harmonic trap, Phys. Rev. A \textbf{95}, 023607 (2017).}

\bibitem{flob15}{J. Flo\ss, A. Kamalov, I. Sh. Averbukh, and P. H. Bucksbaum, Observation of Bloch oscillations in molecular rotation, Phys. Rev. Lett. \textbf{115}, 203002 (2015).}

\bibitem{chalopin20}{T. Chalopin, T. Satoor, A. Evrard, V. Makhalov, J. Dalibard, R. Lopes, and S. Nascimb\`{e}ne, Probing chiral edge dynamics and bulk topology of a synthetic Hall system, Nat. Phys. \textbf{16}, 1017-1021 (2020).}

\bibitem{ozawa19a}{T. Ozawa, H. M. Price, A. Amo, N. Goldman, M. Hafezi, L. Lu, M. C. Rechtsman, D. Schuster, J. Simon, O. Zilberberg, and I. Carusotto, Topological photonics, Rev. Mod. Phys. \textbf{91}, 015006 (2019).}

\bibitem{smirnova20}{D. Smirnova, D. Leykam, Y. Chong, and Y. Kivshar, Nonlinear topological photonics, Appl. Phys. Rev. \textbf{7}, 021306 (2020).}

\bibitem{hafezi13}{M. Hafezi, S. Mittal, J. Fan, A. Migdall, and J. M. Taylor, Imaging topological edge states in silicon photonics, Nat. Photonics, \textbf{7} (12), 1001-1005 (2013).}

\bibitem{mukherjee17}{S. Mukherjee, A. Spracklen, M. Valiente, E. Andersson, P. \"{O}hberg, N. Goldman, and R. R. Thomson, Experimental observation of anomalous topological edge modes in a slowly driven photonic lattice, Nat. Commun. \textbf{8}, 13918 (2017).}

\bibitem{rechtsman13}{M. C. Rechtsman, J. M. Zeuner, Y. Plotnik, Y. Lumer, D. Podolsky, F. Dreisow, S. Nolte, M. Segev, and A. Szameit, Photonic Floquet topological insulators, Nature \textbf{496} (7444), 196-200 (2013).}

\bibitem{mittal16}{S. Mittal, S. Ganeshan, J. Fan, A. Vaezi, M. Hafezi, Measurement of topological invariants in a 2D photonic system, Nat. Photonics \textbf{10} (3), 180 (2016).}

\bibitem{weiwang16}{D.-W. Wang, H. Cai, R.-B. Liu, and M. O. Scully, Mesoscopic superposition states generated by synthetic spin-orbit interaction in Fock-state lattices, Phys. Rev. Lett. \textbf{116}, 220502 (2016).}

\bibitem{cai21}{H. Cai, D.-W. Wang, Topological phases of quantized light, Natl. Sci. Rev. \textbf{8}, nwaa196 (2021).}

\bibitem{saugmann22}{P. Saugmann and J. Larson, A Fock state lattice approach to quantum optics, arXiv:2203.13813 (2022).}

\bibitem{zibold10}{T. Zibold, E. Nicklas, C. Gross, and M. K. Oberthaler, Classical bifurcation at the transition from Rabi to Josephson dynamics, Phys. Rev. Lett. \textbf{105}, 204101 (2010).}

\bibitem{smerzi97}{A. Smerzi, S. Fantoni, S. Giovanazzi, and S. Shenoy, Quantum coherent atomic tunneling between two trapped Bose-Einstein condensates, Phys. Rev. Lett. \textbf{79}, 4950 (1997).}

\bibitem{andrews97}{M. R. Andrews, C. G. Townsend, H.-J. Miesner, D. S. Durfee, D. M. Kurn, and W. Ketterle, Observation of interference between two Bose condensates, Science \textbf{275} 5300, (1997).}

\bibitem{weiss06}{C. Weiss and T. Jinasundera, Coherent control of mesoscopic tunneling in a Bose-Einstein condensate, Phys. Rev. A \textbf{72}, 053626 (2005).}

\bibitem{kidd19}{R. Kidd, M. Olsen, and J. Corney, Quantum chaos in a Bose-Hubbard dimer with modulated tunneling, Phys. Rev. A \textbf{100}, 013625 (2019).}

\bibitem{vera13}{J. Lozada-Vera,V. S. Bagnato, and M. C. de Oliveira, Coherent control of quantum collapse in a bosonic Josephson junction by modulation of scattering length, New J. Phys. \textbf{15}, 113012 (2013).}

\bibitem{engelhardt13}{G. Engelhardt, V. M. Bastidas, C. Emary, and T. Brandes, AC-driven quantum phase transition in the Lipkin-Meshkov,Glick model, Phys. Rev. E \textbf{87}, 052110 (2013).}

\bibitem{watanabe12}{G. Wantanabe and H. M\"{a}kel\"{a}, Floquet analysis of the modulated two-mode Bose-Hubbard model, Phys. Rev. A \textbf{85}, 053624 (2012).}

\bibitem{sorensen05}{A. S. S{\o}rensen, E. Demler, and M. D. Lukin, Fractional quantum Hall states of atoms in optical lattices, Phys. Rev. Lett. \textbf{94}, 086803 (2005).}

\bibitem{wang10}{J. Wang and J. Gong, Generating a fractal butterfly Floquet spectrum in a class of SU(2) systems, Phys. Rev. E \textbf{81}, 026204 (2010).}

\bibitem{chin10}{C. Chin, R. Grimm, P. Julienne, and E. Tiesinga, Feshbach resonances in ultracold gases, Rev. Mod. Phys. \textbf{82}, 1225 (2010).}

\bibitem{abeelen99}{F. A. von Abeelen and B. J. Verhaar, Time-dependent Feshbach resonance scattering and anomalous decay of a Na Bose-Einstein Condensate, Phys. Rev. Lett. \textbf{83}, 1550 (1999).}

\bibitem{thalhammer06}{G. Thalhammer, K. Winkler, F. Lang, S. Schmid, R. Grimm, and J. H. Denschlag, Long-lived Feshbach molecules in a three-dimensional optical lattice, Phys. Rev. Lett. \textbf{96}, 050402 (2006).}

\bibitem{cornish00}{S. L. Cornish, N. R. Claussen, J. L. Roberts, E. A. Cornell, and C. E. Wieman, Stable ${}^{85}\mathrm{Rb}$ Bose-Einstein condensates with widely tunable interactions, Phys. Rev. Lett. \textbf{85}, 1795 (2000).}

\bibitem{thompson05}{S. T. Thompson, E. Hodby, and C. E. Wieman, Ultracold molecule production via a resonant oscillating magnetic field, Phys. Rev. Lett. \textbf{95}, 190404 (2005).}

\bibitem{thalhammer08}{G. Thalhammer, G. Barontini, L. De Sarlo, J. Catani, F. Minardi, and M. Inguscio, Double species Bose-Einstein condensates with tunable interspecies interactions, Phys. Rev. Lett. \textbf{100}, 210402 (2008); C. Weber, G. Barontini, J. Catani, G. Thalhammer, M. Inguscio, and F. Minardi, Association of ultracold double-species bosonic molecules, Phys. Rev. A \textbf{78}, 061601(R) (2008).}

\bibitem{tanzi18}{L. Tanzi, C. R. Cabrera, J. Sanz, P. Cheiney, M. Tomza, and L Tarruell, Feshbach resonance in potassium Bose-Bose mixtures, Phys. Rev. A \textbf{98}, 062712 (2018).}

\bibitem{pollack10}{S. E. Pollack, D. Dries, R. G. Hulet, K. M. F. Magalh\~{a}es, E. A. L. Henn, E. R. F. Ramos, M. A. Caracanhas, and V. S. Bagnato, Collective excitation of a Bose-Einstein condensate by modulation of the atomic scattering length, Phys. Rev. A \textbf{81}, 053627 (2010).}

\bibitem{meinert16}{F. Meinert, M. J. Mark, K. Lauber, A. J. Daley, and H.-C. N\"{a}gerl, Floquet engineering of correlated tunneling in the Bose-Hubbard model with ultracold atoms, Phys. Rev. Lett. \textbf{116}, 205301 (2016).}

\bibitem{greschner14}{S. Greschner, G. Sun, D. Poletti, and L. Santos, Density-dependent synthetic gauge fields using periodically modulated interactions, Phys. Rev. Lett. \textbf{113}, 215303 (2014).}

\bibitem{inouye98}{S. Inouye, M. R. Andrews, J. Stenger, H.-J. Miesner, D. M. Stamper-Kurn, and W. Ketterle, Observation of Feshbach resonances in a Bose-Einstein condensate, Nature \textbf{392} 151-154 (1998).}

\bibitem{bilitewski15}{T. Bilitewski and N. Cooper, Population dynamics in a Floquet realization of the Harper-Hofstadter Hamiltonian, Phys. Rev. A \textbf{91}, 063611 (2015).}

\bibitem{bianco11}{R. Bianco and R. Resta, Mapping topological order in coordinate space, Phys. Rev. B \textbf{84}, 241106(R) (2011).}

\bibitem{buchhold12}{M. Buchhold, D. Cocks, and W. Hofstetter, Effects of smooth boundaries on topological edge modes in optical lattices, Phys. Rev. A \textbf{85}, 063614 (2012).}

\bibitem{klitzing80}{K. v. Klitzing, G. Dorda, and M Pepper, New method for high-accuracy determination of the fine-structure constant based on quantized Hall resistance, Phys. Rev. Lett. \textbf{45}, 494 (1980).}

\bibitem{wang13}{L. Wang, A. Soluyanov, and M Troyer, Proposal for direct measurement of topological invariants in optical lattices, Phys. Rev. Lett. \textbf{110}, 166802 (2013).}

\bibitem{wang16}{F. Wang, X. Li, D. Xiong, and D. Wang, A double species ${}^{23}$Na and ${}^{87}$Rb Bose-Einstein condensate with tunable miscibility via interspecies resonance, J. Phys. B: At. Mol. Opt. Phys. \textbf{49}, 015302 (2016).}

\bibitem{lee18}{K. L. Lee, N. B. J{\o}rgensen, L. J. Wacker, M. G. Skou, K. T. Skalmstrang, J. J. Arlt, and N. P. Proukakis, Time-of-flight expansion of binary Bose-Einstein condensates at finite temperature, New J. Phys. \textbf{20}, 053004 (2018).}

\bibitem{kolovsky14}{A. R. Kolovsky, F. Grusdt, and M. Fleischhauer, Quantum particle in a parabolic lattice in the presence of a gauge field, Phys. Rev. A \textbf{89}, 033607 (2014).}

\bibitem{wang21}{W.-Y. Wang, J. Lin, and J. Liu, Cyclotron dynamics of a Bose-Einstein condensate in a quadruple-well potential with synthetic gauge fields, Front. Phys. \textbf{16}, 52502 (2021).}



\end{thebibliography}
\end{document}